\documentclass[%
reprint,
 amsmath,amssymb,
 aps,
]{revtex4-2}
\usepackage{float}
\makeatletter
\let\newfloat\newfloat@ltx
\makeatother

\usepackage{graphicx}
\usepackage{dcolumn}
\usepackage{algorithm}
\usepackage{algorithmic}

\usepackage{appendix}
\usepackage{color}
\usepackage{xcolor}
\usepackage{bm}

\begin{document}


\title{Tuning the kinetics of intracellular transport}

\author{Ardra Suchitran}
\email{ardra235@gmail.com}
 \affiliation{Department of Physics, Government College for Women - Thiruvananthapuram, Kerala 695014, India}
\author{Sreekanth K Manikandan}%
 \email{sreekm@stanford.edu}
\affiliation{Department of Chemistry, Stanford University, Stanford, CA, USA 94305
}%

\date{\today}

\begin{abstract}
A variety of complex mechanisms, from chemical reaction pathways to active fluctuations, orchestrate molecular transport in intracellular environments. Despite significant recent progress in visualizing and probing these processes, little is known about how tunable the resulting dynamics is through external physical controllers. Here, we demonstrate that coarse-grained, reinforcement learning-based protocols can be developed to achieve highly localized and targeted cargo transport by kinesin motors on intracellular tracks. These protocols can be implemented in practice using optical tweezers, and their feasibility is showcased within experimentally relevant parameter regimes. Our results open new avenues for targeted control of intracellular transport processes, especially when opportunities for control are limited.
\end{abstract}

\maketitle

\begin{figure*}
    \centering
    \includegraphics[width=0.8\linewidth]{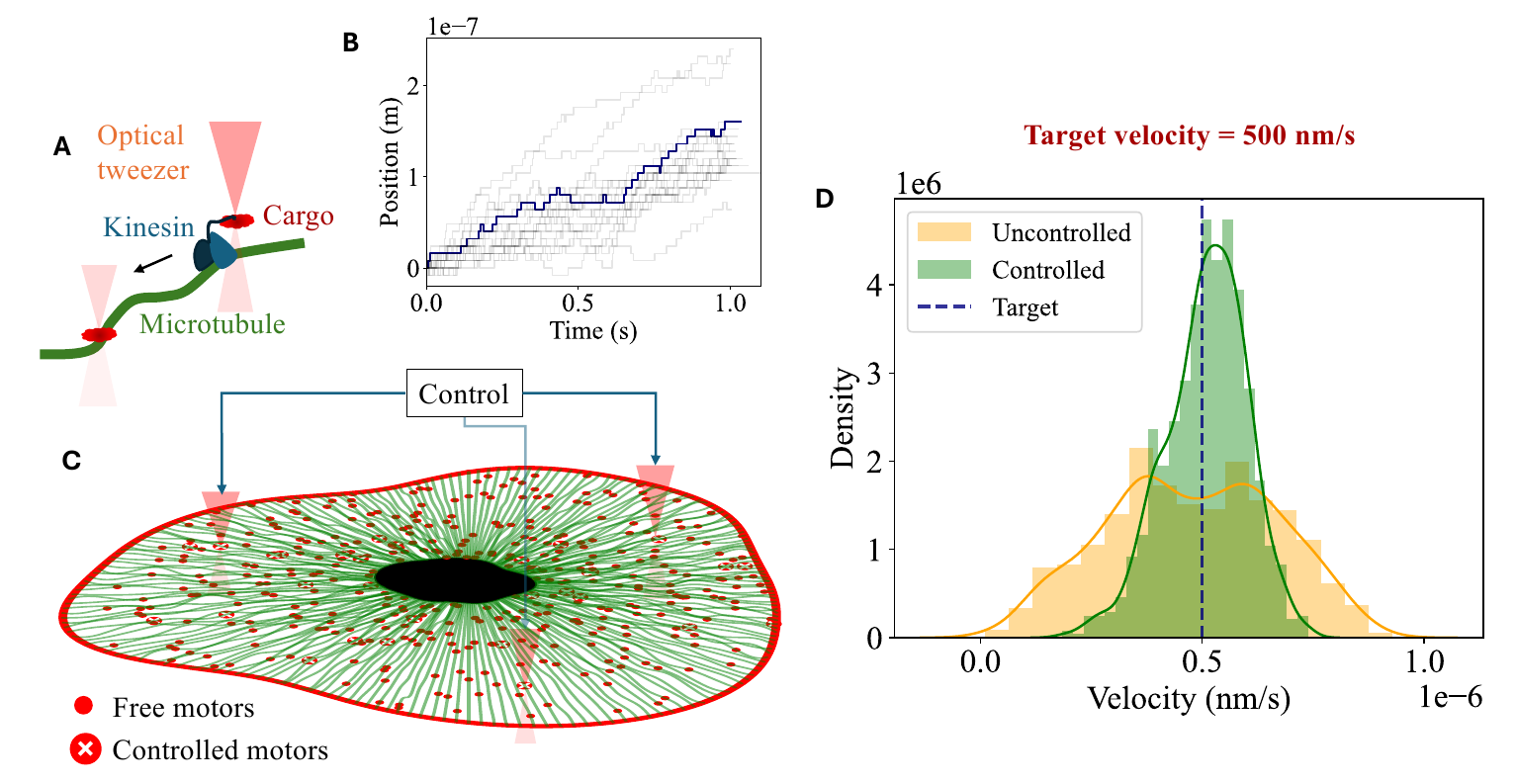}
    \caption{(A) Schematic representation of a cargo-carrying kinesin walking along a microtubule track, with the cargo acting as a hindering load that is probed by an optical tweezer. (B) Sample trajectories of kinesin stepping along the microtubule, illustrating both forward steps and occasional reverse steps. (C) Two-dimensional model of kinesin-mediated transport, showing free and controlled kinesins distributed along microtubule tracks. (D) Distributions of velocities of uncontrolled and controlled motors, obtained from 1000 samples.}
    \label{fig:Schematics}
\end{figure*}

Intracellular transport is a fundamental process in cellular biophysics, responsible for the movement of various cargos such as organelles, vesicles, and proteins within a cell \cite{mogre2020getting,iversen2011endocytosis}. A significant fraction of this transport is facilitated by molecular motors \cite{schliwa2003molecular, vale2003molecular}, such as kinesins \cite{hirokawa2009kinesin} and dyneins \cite{roberts2013functions}, that move along cellular tracks with specific polarity \cite{welte2004bidirectional}. The efficiency and precision of these processes are crucial for maintaining a wide range of cellular functions, including nutrient distribution \cite{hesketh2002intracellular}, signal transduction \cite{kholodenko2003four}, cell division \cite{sawin1991motor}, development \cite{supp2000molecular}, and cellular homeostasis \cite{ballabio2020lysosomes,vigani2019essential}.

Experimental studies over the last few decades have demonstrated that cargo proteins, such as kinesin-1, move along microtubules through a stepping mechanism: kinesin takes 8 nm steps with each ATP hydrolysis \cite{schnitzer1997kinesin}. This movement involves alternating the two catalytic heads in a "hand-over-hand" fashion, resembling bipedal walking \cite{yildiz2004kinesin}. From this mechanistic picture, a chemical reaction network model for molecular transport emerges, where each step of the motor protein is represented as a series of chemical reactions \cite{clancy2011universal,fisher2001simple,coppin1997load}. This framework has since proven to be extremely useful for analyzing how external loads and mechanical stresses influence transport dynamics \cite{carter2005mechanics,block2007kinesin,block1990bead}. 
For instance, the load dependence of the stepping kinetics of kinesin has been tested in minimal reconstituted experiments, and is found to be well-described by the Bell's equation \cite{bell1978models,ariga1}: 
\begin{equation}
\label{eq:rate}
k(F) = k_0 \exp\left(\frac{F d}{k_B T}\right),
\end{equation}
which provides insight into how an external force \( F \) affects the rate constant \( k\) of a reaction. Here \( k_0 \) is the rate constant in the absence of any external force, \( d \) is a length scale which can be obtained as a fitting parameter from experimental data \cite{ariga1,ariga2}, \( k_B \) is the Boltzmann constant, and \( T \) is the temperature. More recent experimental and theoretical studies have further shown that kinesin also accelerates under intracellular noise \cite{ariga2,feng2023unraveling}, and hence, perhaps, is evolutionarily designed to perform well in complex environments \cite{vale2000way}. Indeed, an unresolved problem is that the velocities of vesicles transported by kinesin in cells are much faster than those observed in vitro \cite{ross2016dark}.

These experiments indicate that external forces and perturbations can in principle be used as an effective means to control and target specific transport properties by molecular motors such as kinesins. 
However, achieving a reasonable degree of control in complex intracellular environments presents several practical challenges. 
First and foremost, opportunities for control in these systems are often fundamentally or practically limited \cite{chennakesavalu2021probing,chennakesavalu2024adaptive} — the environment is highly fluctuating, and there are often only a few, and sometimes just a single degree of freedom that can be probed. Second, the complexity of underlying biochemical pathways \cite{rothman1994mechanisms} can be a significant barrier, as detailed knowledge is either unknown or too complex for deciphering practical controls. Additionally, available physical controllers such as optical tweezers are restricted by their maximum force exertion and spatiotemporal resolution \cite{bradac2018nanoscale}. Effective solutions should thus involve designing controls that are appropriately coarse-grained. Finally, implementing controls in a minimally invasive manner is crucial to avoid disrupting the biological system of interest \cite{zhang2021plasmonic,norregaard2017manipulation}.

In this work, we present a computational framework which addresses a number of these open challenges. We first develop a two-dimensional model of kinesin-mediated transport, which takes into account the effects of variable cargo sizes, spatially inhomogenous intracellular noise, the underlying microtubule network architecture and kinesin - kinesin interactions. By integrating this model with reinforcement learning strategies \cite{mnih2013playing}, we demonstrate that targeted and highly localized control of cargo transport can be achieved in a dynamic fashion (See Fig.\ \ref{fig:Schematics}A - D). We use experimental parameters for the models and controls, and the resulting protocols are chosen to be realized in optical tweezers \cite{moffitt2008recent,curtis2002dynamic}. Our results open new avenues for targeted control of intracellular transport processes, especially when opportunities for control are limited.

\begin{figure*}
    \centering
    \includegraphics[width=0.8\linewidth]{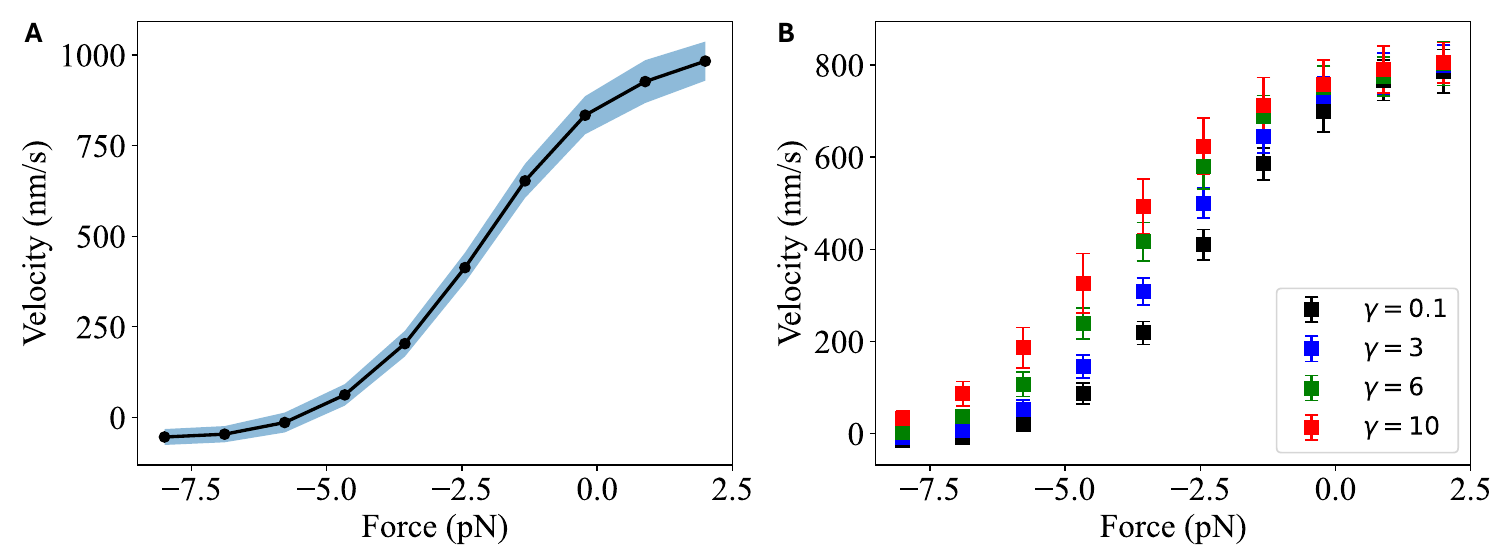}
    \caption{(A) This figure illustrates the dependence of kinesin velocity on the applied load force at a fixed ATP concentration. The results are in excellent agreement with the experimental and numerical data from Ariga et al. \cite{ariga1,ariga2}, accurately reproducing the quantitative relationship between kinesin velocity and load force. Note that the velocity reduces to zero at the stalling forces of the motor, which is around -6 {\it pN}. (B) This figure demonstrates the dependence of kinesin velocity on the strength of intracellular noise, in good agreement with previous studies. Here, $\gamma$ is the scaling parameter of the Lévy noise, where higher values of $\gamma$ indicate greater variance in the noise. The simulations use noise distributions truncated at a physiologically plausible value of 30 {\it pN}. The results show that kinesin velocity increases with the application of noise, consistent with the noise-induced acceleration observed in earlier research. The parameter values are chosen from Refs. \cite{ariga1,ariga2} and are given by: \( k_2 = 981 \), \( k_3 = 22.8 \), \( k_1 = 129 \), \( t_{\text{max}} = 1 \), \( \Gamma = 3.1 \times 10^{-8} \), \( k = 0.075 \times 10^{-3} \), \( d_f = 3.3 \times 10^{-9} \), \( d_b = 0.47 \times 10^{-9} \), \( k_b = 1.38 \times 10^{-23} \), and \( T = 300 \).}
    \label{fig:benchmark}
\end{figure*}

\textit{Results}: We first develop a two-dimensional model of kinesin-mediated transport in complex intracellular settings. This model features a network of microtubule tracks radiating from the center of the cell outward in wiggly patterns, reflecting the radial organization typical of real cells. A fixed number of motors carrying cargoes are attached to these tracks at any given point in time. Each motor is connected to the cargo (which in some cases is a probe particle) via a linear spring with stiffness $K$. At each time point, the motor can either step forward or backward with a step size $d$ of 8 nm. This movement of kinesin is described by a two-state kinetic model with two internal states, as first introduced in \cite{ariga1}. There are three transition paths between these states. The first is a load-independent path (state 1 to state 2) with a rate constant $k_c$, corresponding to an internal transition. The second and third paths are load-dependent (state 2 to state 1) with rate constants $k_f$ and $k_b$, corresponding to mechanical transitions for forward and backward steps, respectively. The load-dependent rate constants are given by Bell's equation \cite{bell1978models,ariga1}:
\begin{align}
    k_{\lbrace{f,b \rbrace}}(F) &= k_{\lbrace{f,b \rbrace}}^0 \exp \left( \frac{d_{\lbrace{f,b \rbrace}} F}{k_B T} \right),
\end{align}
where $F$ is an external constant force called load, $k_{\lbrace{f,b \rbrace}}^0$ are the rate constants at zero load, $d_{\lbrace{f,b \rbrace}}$ are characteristic distances for forward and backward steps, $k_B$ is the Boltzmann constant, and $T$ is the absolute temperature.

The dynamics of the probe, which is pulled by the kinesin motor, are described by the Langevin equation:
\begin{align}
\label{eq:probe}
    \Gamma\frac{d}{dt}{\bm x}_p &= K({\bm x}_m - {\bm x}_p) + {\bm F}_p ({\bm x}_p) + {\bm \xi},
\end{align}
where $\Gamma$ is the viscous drag, $K$ is the stiffness of the link connecting the probe to the motor, ${\bm x}_m$ are the positions of the motor, ${\bm x}_p$ are the positions of the probe and ${\bm \xi}$ represents white Gaussian thermal fluctuations that satisfy $\langle \xi_i(t) \rangle = 0$ and $\langle \xi_i(t) \xi_j(s) \rangle = 2k_BT\Gamma\delta_{ij}\delta(t-s)$. The remaining force term ${\bm F}_p = {\bm F}_0 + {\bm F}_n + {\bm F}_q$, where ${\bm F}_0$ is the inherent load force from the cargo, which we assume to be a variable one from motor to motor. The term ${\bm F}_n$ is a fluctuating force with zero mean, representing intracellular active noise. We further assume that ${\bm F}_n$ depends nontrivially on ${\bm x}$ as discussed later. The third term ${\bm F}_q$ corresponds to interactions between the motors, accounting for the experimentally observed jamming and crowding effects \cite{leduc2012molecular,reese2011crowding,klumpp2019life,lakadamyali2014navigating}, which essentially slow down motors. In this work, we employ this force contribution through a method commonly referred to as quorum sensing \cite{duan2023dynamical,miller2001quorum,bauerle2018self}. Specifically, if a motor enters a crowded neighborhood, all the motors (probes) in that region experience a hindering force, which is close in magnitude to the stalling force of the motors.

In the simulations, we first specify an outer irregular periphery for a cell and an inner region corresponding to the nucleus (filled black region in Fig.\ \ref{fig:Schematics}C), and a collection of wiggly microtubule tracks connecting the two. All the kinesins are initially bound to microtubules and carry a randomly chosen cargo of load $\in [0, 1, 2, 3]$  {\it pN}. We assume that some of these cargoes are probes which can be manipulated through optical tweezers \cite{beeg2008transport}. The interior of the cell is assumed to be an infinite reservoir for more kinesins and cargoes. The time evolution of the system is implemented in two steps: Step 1 - All kinesins take a step through chemical reactions and move along the microtubules, and Step 2 - the position of the cargo/probe evolves through Langevin dynamics to the new location of the motor. This process continues cyclically, driving the time evolution of the entire system. Whenever a motor reaches the periphery, it is assumed to be detached from the microtubule, and a new motor and cargo attaches at a position randomly chosen along the length of the microtubule.

This model, without the interactions between the motors, variable loads and spatially inhomogenous intracellular noise, is statistically identical to the one-dimensional model considered in Refs.\ \cite{ariga1,ariga2}, and reproduce their results in the respective limits. In Fig. 2A, we show the dependence of kinesin velocity on the load force at a fixed ATP concentration. Note that the velocity reduces to zero at the stalling forces of the motor, which is around $-6$ {\it pN}. In Fig. 2B, we show the dependence of kinesin velocity on the strength of the intracellular noise, which is modelled as Levy noise as discussed in Ref.\ \cite{ariga2}. Here, $\gamma$ is the scaling parameter of the Levy noise. The higher the value of $\gamma$, the higher the variance of the noise. Furthermore, the simulations are conducted with noise distributions that are truncated at a physiologically plausible value of 30 {\it pN}. We find that the velocity of kinesin increases under the application of the noise, consistent with the noise-induced acceleration observed in previous studies \cite{ariga2,feng2023unraveling}. The theoretical understanding of this observation follows from the application of Jensen's inequality to Eq.\ \eqref{eq:rate}. Leveraging the convexity of the exponential function, it can be shown that the expected rate constant under fluctuating forces satisfies:
\begin{equation}
\langle k(F) \rangle > k\left(\langle F \rangle\right).
\end{equation}
This inequality indicates that the average rate constant generally increases when the driving forces fluctuate. In the case of kinesin, due to different length scales $d$ for forward versus reverse reactions, the enhancement of the forward stepping rate for a given strength of fluctuations is greater than that for the reverse. This results in the acceleration of the motors \cite{ariga2}. Based on the
universality of Jensen's inequality and Eq.\ \eqref{eq:rate},
Ref.\ \cite{ariga2} further argues that any enzyme obeying the same rate equation
can experience noise-induced acceleration. As they point out, the observed enhancement in motor velocity is not dependent on the type of noise. Hence in the following, for computational simplicity, we account for the noise induced effect by simply re-scaling the temperature in Eq.\ \eqref{eq:probe}, with a scaling factor ${\bm A} ({\bm x})$.

{\it Tuning the kinetics of kinesin transport}: As discussed, the in-vivo kinetics of molecular motors and their transport velocity are significantly affected by variable hindering loads, intracellular noise, the nature of the underlying transport network, and intracellular crowding effects. Assuming that the motors can be probed through optical forces, identifying spatiotemporal control forces that can lead to targeted transport properties is a highly non-trivial optimization problem. Interestingly, reinforcement learning (RL) has emerged as a promising approach for tackling such challenges. In an RL setting, the control problem is formulated as the optimization of a Markov decision process (MDP) \cite{puterman1990markov}, enabling the use of deep reinforcement learning algorithms for solving it. The complexity and high dimensionality of these problems have been addressed by significant advancements in deep learning, facilitating the optimization of complex high-dimensional feedback protocols. Recently, the feasibility of this approach was demonstrated in a biophysical context through the assembly of branched-actin networks with spatially patterned structural properties and nontrivial response properties \cite{chennakesavalu2024adaptive}. Here, we use reinforcement learning to identify the control forces that will lead to the transport of interacting kinesin motors with targeted velocities.

As control forces, we consider a discrete set of external loads ranging from -7 {\it pN} to 4 {\it pN} in steps of 1 {\it pN}. This is to address potential experimental challenges associated with applying load forces with arbitrary resolution. We then train RL agents that can modulate the external load in a feedback fashion to achieve desired kinesin velocities. Training involves running the dynamics of kinesins for each episode of duration $t_{max} = 1s$. For each episode, the current radial position of motors ${\bm r}({\bm x})$, velocity ${\bm v}({\bm x})$, noise amplitude on the probe ${\bm A}({\bm x})$ (An integer value between 1 and 10), inherent cargo load on the motor ${\bm L}({\bm x})$ (randomly chosen value between 0 and -3 {\it pN}), and quorum sensing state ${\bm Q}({\bm x})$ are measured, and a corresponding action is taken. This process is repeated until a batch of experiences are collected. The agent is then trained to reduce the error between the achieved velocity and the target velocity, using a policy gradient approach \cite{SpinningUp2018} - see Algorithm.\ \ref{alg:vpg_kinesin}.

\begin{algorithm}
\caption{ }
\label{alg:vpg_kinesin}

\begin{algorithmic}[0]
\STATE \textbf{Input:} Action Space ${\bm a}  \in \{-7, -6, \ldots, 4\}$ {\it pN}, episodes $N$, learning rate $\alpha$, replay buffer size $B$
\STATE \textbf{Initialize:}
\STATE ${\bm L}$: Inherent cargo load $\in \{0, -1, -2, -3\}$ {\it pN}
\STATE ${\bm Q}({\bm x})$: Quorum sensing state $\in \{0, 1\}$
\STATE ${\bm A}({\bm x})$: Noise amplitude $\in \{1, \ldots, 10\}$
\STATE Initialize policy network $\pi_\theta$ and optimizer
\STATE Initialize replay buffer $\mathcal{R}$
\STATE Initialize, cell background, microtubule tracks, and motors with cargo
\FOR{episode $e = 1$ \textbf{to} $N$} 
    \STATE Measure state: ${\bm s} = [{\bm r}({\bm x}), {\bm v}({\bm x}), {\bm A}({\bm x}), {\bm L}, {\bm Q}]$
    \STATE Select action ${\bm a}$ using policy $\pi(s)$
    \STATE Apply action ${\bm a}$, update state
    \STATE Measure new position ${\bm r}$ and velocity ${\bm v}$
    \STATE Calculate reward: $R = -\text{MSE}({\bm v}({\bm{x}}), {\bm v}_{target}(\bm{x}))$
    \STATE Store transition $({\bm s}, {\bm a}, R, {\bm s}')$ in $\mathcal{R}$
    \IF{size of $\mathcal{R}>B$}
        \STATE Sample batch from $\mathcal{R}$
        \STATE Perform training step
    \ENDIF
\ENDFOR

\STATE \textbf{Output:} Trained policy $\pi_\theta$
\end{algorithmic}
\end{algorithm}

Remarkably, this approach enables us to target specific kinesin velocities with high fidelity, even in the presence of significant inhomogeneity in the state space. In Figure 3A, we show the radial trajectories of controlled motors (green) versus uncontrolled motors (red). Here, the target velocity is set to 500 nm/s.  The distribution over an ensemble of such motors is shown in Fig.\ \ref{fig:Schematics}D. Note that the controlled motors all have velocities close to the target, while the uncontrolled ones significantly deviate, featuring both higher and lower velocities than the target value. 

\begin{figure}
    \centering
    \includegraphics[width=.99\linewidth]{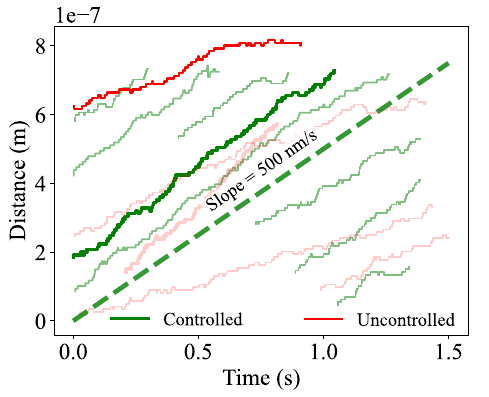}
    \caption{(A) Radial trajectories of controlled motors (green) versus uncontrolled motors (red), with the target velocity set to 500 nm/s. The motors are initialized at different locations in the cell, resulting in different initial values for the distance. The distribution over ensembles of such motors was previously shown in Figure 1D. Controlled motors exhibit velocities close to the target, while uncontrolled motors show significant deviations, featuring both higher and lower velocities than the target. The parameter values used are taken from Ref.\ \cite{ariga1}, and are: \( k_2 = 981 \), \( k_3 = 22.8 \), \( k_1 = 129 \), \( t_{\text{max}} = 1 \), \( \Gamma = 3.1 \times 10^{-8} \), \( k = 0.075 \times 10^{-3} \), \( d_f = 3.3 \times 10^{-9} \), \( d_b = 0.47 \times 10^{-9} \), \( k_b = 1.38 \times 10^{-23} \), and \( T = 300 \).}
    \label{fig:benchmark}
\end{figure}

The protocols identified in this way to achieve targeted kinesin velocities are nondeterministic: depending on the current state of the system, different loads are required to maintain a particular velocity and to transition between different velocities. In Figures 4A - 4C, we show a heatmap corresponding to the correlations between three of the state variables — the free cargo load on the motor, the quorum sensing state and the noise scale — with the chosen control force. Even though the action space contains control forces in the range -7 to 4 {\it pN}, the controller essentially switches between -1 to 1 {\it pN} for the chosen target velocity.  For higher free cargo loads and active quorum sensing, the motor velocities are expected to reduce, and the controller is found to predominantly employ +1 {\it pN} force which has a pushing effect on the motor. At higher noise amplitudes, where the motors are expected to accelerate, the agent predominantly applies a hindering load of  -1 {\it pN}. 

Note that deriving an equivalent deterministic dependence at the level of individual motors is challenging, especially due to the strong dependencies of the chosen action on the other state variables and the overall stochastic nature of the problem. Alternatively, RL seems to effectively handle these complexities by leveraging the stochasticity and the multi-dimensional state space to find optimal control policies. This highlights the potential of our approach in dealing with highly dynamic and noisy intracellular environments where traditional deterministic or pre-designed strategies might fall short.

\begin{figure}
    \centering
    \includegraphics[width=\linewidth]{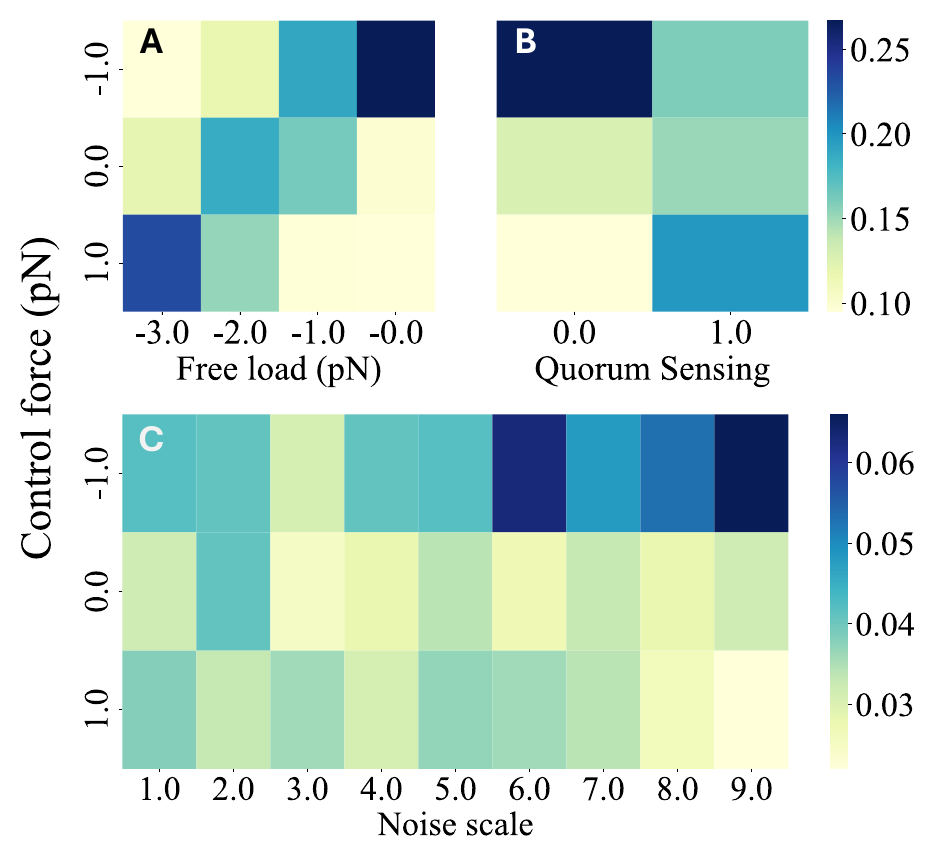}
    \caption{Heatmaps showing the correlations between two state variables — (A) the free cargo load on the motor, (B) the quorum sensing state and (C) The noise scale —with the chosen control force. Note that, even though the action space contains control forces in the range -7 to 4 {\it pN}, the controller essentially switches between -1 to 1 {\it pN} for the chosen target velocity. For higher free cargo loads and active quorum sensing, the motor velocities are expected to reduce, and the controller is found to predominantly employ +1 {\it pN} force which has a pushing effect on the motor. At higher noise amplitudes, where the motors are expected to accelerate, the agent predominantly applies a hindering load of  -1 {\it pN}. }
    \label{fig:benchmark}
\end{figure}

For the illustrations, we have assumed that all, if not a subset of motors are individually controllable. Control at the level of a subset of motors can indeed be achieved using appropriately engineered cargoes \cite{sims2009probing,beeg2008transport} and holographic optical tweezers \cite{kirkham2015precision,mas2014force}. However, individually controlling all motors is highly challenging in practice. We would like to point out that this is not necessarily a drawback for the methods we propose. Depending on experimental feasibility, one could design spatially inhomogeneous control schemes and use our methods to identify effective control policies \cite{chennakesavalu2021probing}. For instance, applying control forces that vary spatially within the cell could still enable targeted manipulation of motor velocities. Moreover, the models we train are highly versatile and capable of simultaneously driving different motors at various locations with different velocities in a multi-agent setting \cite{busoniu2008comprehensive,tan1993multi}. By leveraging this, our methods can potentially target collective behaviors of motor proteins \cite{campas2006collective,celis2015correlations}, leading to more efficient and coordinated intracellular transport even when individual control is impractical. 


In summary, we have demonstrated that locally coarse-grained, reinforcement learning-based controls can be developed to achieve tunable and targeted cargo transport by interacting kinesin motors on intracellular tracks. Both the model and controls function within experimentally relevant parameter regimes and conditions that mimic intracellular states. They include the architecture of intracellular tracks, spatially inhomogeneous noise distributions, and crowding effects due to kinesins' proximity to each other. Despite its computational simplicity, the model captures experimentally verified features such as the acceleration of motors under intracellular noise. The optimal control protocols derived are non-deterministic and local, and designed for practical implementation in experiments. It will be interesting to see whether the observed degree of control persists in both {\it in-vitro} and {\it in-vivo} experiments.

Our model can be further developed by incorporating the dynamics of microtubules, interactions at microtubule junctions, and the effects of other molecular motors such as myosins and dyneins. On this regard, studying bidirectional cargo transport and the "tug-of-war" phenomenon \cite{welte2004bidirectional}, along with their controllability, could be an interesting follow-up work. Additionally, exploring kinesins on microtubules as an active matter system could provide insights into the collective and emergent behaviours \cite{ramaswamy2010mechanics} in these systems. 
\section*{Code Availability}
See \url{https://github.com/sreekmnoneq/Kinesin-RL} for a basic implementation of the results in the paper.

\section*{Acknowledgements}
Authors thank Takayugi Ariga for helpful discussions on Refs. \cite{ariga1,ariga2}. Authors thank the Kerala Theoretical Physics Initiative - Active Research Training (KTPI - ART) program for facilitating the research collaboration. 
SKM acknowledges the Knut and Alice Wallenberg Foundation for financial support through Grant No. KAW 2021.0328. SKM thanks Biswajit Das, Light Matter Lab, IISER Kolkata, India for helpful discussions on optical trap experiments. 

%

\end{document}